\documentclass[10pt,journal,compsoc]{IEEEtran}
\newif\ifpeerreview

\peerreviewtrue

\usepackage[nocompress]{cite}
\usepackage{url}
\usepackage{amsmath,amssymb,graphicx}
\usepackage{lipsum}
\usepackage[switch]{lineno}
\usepackage{xcolor}

\usepackage{subcaption} 

\usepackage{tikz}
\usetikzlibrary{arrows.meta, positioning, calc, fit, backgrounds}

\usepackage{svg}
\usepackage{booktabs}
\usepackage{multirow}

\usepackage{xcolor}    % For text colors
\usepackage{lipsum}    % Lorem Ipsum generator

\usepackage{siunitx}

\usepackage{glossaries}
\glsdisablehyper

\makeglossaries
\newacronym{psnr}{PSNR}{peak signal-to-noise ratio}
\newacronym{swir}{SWIR}{shortwave infrared}
\newacronym{gsst}{GSST}{Ge$_2$Sb$_2$Se$_4$Te$_1$}
\newacronym{sam}{SAM}{spectral angle mapper}
\newacronym{mse}{MSE}{mean squared error}
\newacronym{aviris}{AVIRIS}{Airborne Visible/Infrared Imaging Spectrometer}
\newacronym{pcm}{PCM}{Phase-Change Material}
\newacronym{fdtd}{FDTD}{Finite-Difference Time-Domain}
\newacronym{rcwa}{RCWA}{Rigorous Coupled-Wave Analysis}
\newacronym{dmd}{DMD}{Digital Micromirror Device}
\newacronym{cs}{CS}{compressive sensing}
\newacronym{em}{EM}{electromagnetic}
\newacronym{snr}{SNR}{signal-to-noise ratio}
\newacronym{Q1}{Q1}{first quadrant}

\makeatletter
\newif\iflipsumenabled
\lipsumenabledfalse

\let\origlipsum\lipsum
\renewcommand{\lipsum}[1][]{%
  \iflipsumenabled
    \textcolor{gray}{\origlipsum[#1]} % Show gray lipsum if enabled
  \else
  \fi
}
\makeatother
\title{\fontsize{22}{22}\selectfont Inverse Design of Metasurface for Spectral Imaging}

\author{Rongzhou~Chen\textsuperscript{*},~Haitao~Nie\textsuperscript{*},~Shuo~Zhu,~Yaping~Zhao,~Chutian~Wang,~Edmund~Y.~Lam%
\IEEEcompsocitemizethanks{\IEEEcompsocthanksitem 
R. Chen, H. Nie, S. Zhu, Y. Zhao, C. Wang, and E. Y. Lam are with the Department of Electrical and Electronic Engineering, The University of Hong Kong, Hong Kong SAR, China.\protect\\
Corresponding author: Edmund Y. Lam (elam@eee.hku.hk).\protect\\
\textsuperscript{*}Co–first authors.}}
\IEEEtitleabstractindextext{%
\begin{abstract}
Inverse design of metasurfaces for the joint optimization of optical modulation and algorithmic decoding in computational optics presents significant challenges, especially in applications such as hyperspectral imaging. We introduce a physics-data co-driven framework for designing reconfigurable metasurfaces fabricated from the phase-change material Ge$_2$Sb$_2$Se$_4$Te$_1$ to achieve compact, compressive spectral imaging in the shortwave infrared region. Central to our approach is a differentiable neural simulator, trained on over $320,000$ simulated geometries, that accurately predicts spectral responses across $11$ crystallization states. This differentiability enables end-to-end joint optimization of the metasurface geometry, its spectral encoding function, and a deep reconstruction network. We also propose a soft shape regularization technique that preserves manufacturability during gradient-based updates. Experiments show that our optimized system improves reconstruction fidelity by up to \SI{7.6}{dB} in the peak-signal-to-noise ratio, with enhanced noise resilience and improved measurement matrix conditioning, underscoring the potential of our approach for high-performance hyperspectral imaging.
\end{abstract}

}

\begin{document}

% \ifpeerreview
% \linenumbers \linenumbersep 15pt\relax
% \fi

\maketitle

% For peer review, we produce line numbers; for camera ready, comment out the \linenumbers block
% \ifpeerreview
% \linenumbers
% \fi

%%%%%%%%%%%%%%%%%%%%%%%%%%%%%%%%%%%%%%%%%%%%%%%%%%%%%%%%%%%%%%%%%%%%%%%%
% 
\IEEEPARstart{H}{yperspectral} imaging acquires hundreds of contiguous spectral bands across a continuous wavelength range, revealing material compositions, with the \gls{swir} region ($1$--\SI{2.5}{\um}) providing significant analytical value through its molecular vibrational signatures and superior penetration properties~\cite{gehm2007single, arnob2018compressed, correa2015snapshot}. By leveraging spectral fingerprints from molecular vibrational overtones and benefiting from low scattering effects, it is highly effective in applications such as biomedical imaging, agriculture monitoring and low-light surveillance. Traditional \gls{swir} hyperspectral imaging setups typically rely on dispersive elements like prisms or gratings~\cite{schaepman2015advanced, guanter2015enmap, rast2019earth}, narrow-band tunable filters~\cite{wang2012liquid, archibald1999development, gat2000imaging}, and Fourier transform interferometers~\cite{garini2006spectral, harvey2004birefringent}, making them bulky, costly, and sometimes suffer from low optical energy efficiency. Alternative methods utilizing single-pixel detectors enhance optical efficiency~\cite{arnob2018compressed, tao2022shortwave}, yet these approaches necessitate precise alignments, challenge field deployment and system miniaturization, and exhibit temporal limitations constraining dynamic scene acquisition. Consequently, significant demand exists for lightweight, compact hyperspectral imaging devices that maintain high optical throughput while addressing these operational constraints.

To address these constraints in traditional hyperspectral imaging systems, recent advances in \gls{pcm} technology present a promising pathway through compact, rapidly reconfigurable metasurface designs~\cite{wuttig2017phase}. The key design challenge is to determine the optimal geometric configurations for the metasurface. Each shape produces unique spectral response characteristics that are essential for effective hyperspectral compressive imaging and enable accurate recovery through decoding algorithms. Among the various phase-change materials available, \gls{gsst} emerges as particularly advantageous due to its exceptional transparency window across the \gls{swir} spectrum and substantial refractive index difference between amorphous and crystalline states~\cite{burtsev2022physical, tao2022shortwave}. This material's adaptability enables continuously variable optical constants through partial crystallization of individual \gls{gsst} elements, forming the foundation for programmable metasurfaces with application-specific spectral responses~\cite{ding2019dynamic, zhang2021reconfigurable}. The inverse design problem thus becomes finding optimal geometric configurations through sophisticated optimization techniques that yield optimal spectral responses for hyperspectral encoding, enabling miniaturized imaging systems without sacrificing performance or requiring bulky optical components.

The computational complexity of inverse design presents challenges for compressive spectral imaging. Traditional approaches using particle swarm optimization or genetic algorithms suffer from slow convergence, high computational costs, and inability to integrate with gradient-based frameworks. By contrast, our approach optimizes only the \gls{gsst} metasurface geometry, leveraging a fixed set of 11 crystallization states to create a set of tunable optical filters that encode high-dimensional spectral information into reduced measurements. The key innovation is our differentiable neural simulator, which accurately approximates electromagnetic responses while enabling end-to-end gradient-based optimization—something impossible with conventional \gls{fdtd} and \gls{rcwa} methods~\cite{amirjani2021computational}. This differentiability facilitates seamless integration of optical encoding and computational decoding~\cite{wiecha2021deep, so2020deep}, while our specialized regularization techniques ensure physical realizability of the optimized designs. Our neural simulator provides the critical bridge between optics and computational reconstruction, enabling optimization that would otherwise require significant computational resources when using traditional electromagnetic solvers. This approach not only improves reconstruction fidelity but also enhances robustness to measurement noise and leads to better-conditioned measurement matrices, overcoming computational bottlenecks that have traditionally hindered joint optimization of metasurfaces with reconstruction networks~\cite{pestourie2020active}.

\begin{figure*}[!t]
\centering
\includegraphics[width=0.85\linewidth]{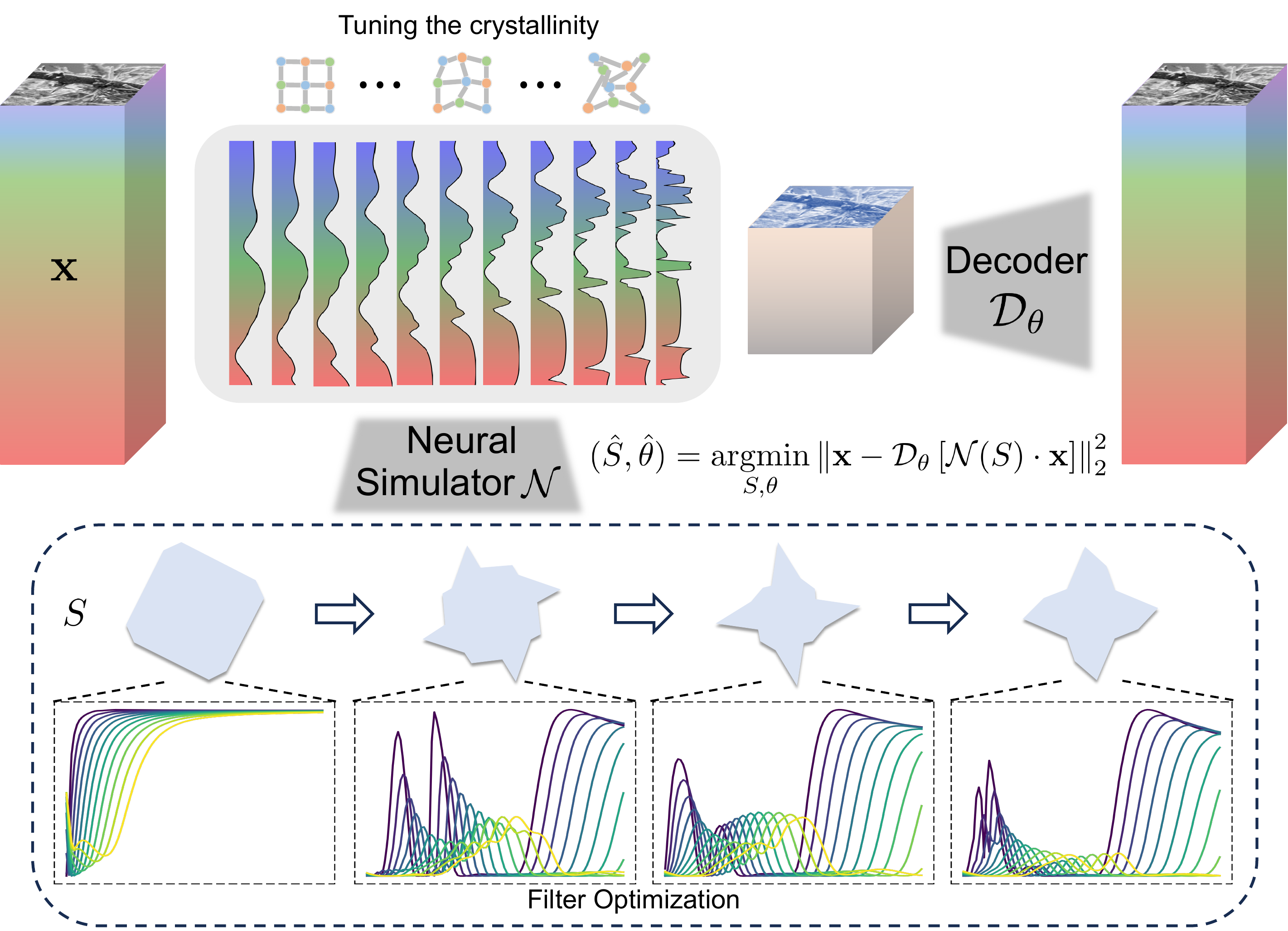}
% \includesvg{figs/cover_v3.svg}
\caption{Illustration of the proposed deep optics framework for compressive hyperspectral imaging using a reconfigurable \gls{gsst} metasurface. By tuning the crystallization state of the metasurface, a diverse set of spectral response functions is generated to serve as tunable optical filters that encode high-dimensional data. A neural simulator, denoted by $\mathcal{N}$, predicts the metasurface’s spectral response based on its shape $S$, enabling an end-to-end, gradient-based optimization of both the optical encoder and the deep reconstruction decoder $\mathcal{D}_\theta$. The diagram also highlights how the filter shapes and their corresponding transmittance spectra evolve during training, emphasizing the benefits of the joint encoder–decoder design in improving overall reconstruction fidelity.}
\label{fig:deep_optics_framework}
\end{figure*}

% \newpage

% \subsection{Contributions} 
In this paper, we solve the \textit{inverse-design} problem for metasurfaces, optimizing GSST structures for compressive spectral imaging. Our framework, shown in Fig.~\ref{fig:deep_optics_framework}, enables gradient optimization by combining a neural simulator with a deep reconstruction network. Unlike previous approaches that rely on non-differentiable optimization~\cite{kennedy1995particle} or intuition-driven design~\cite{lalau2013adjoint}, our approach provides performance improvements. Our main contributions are:
\begin{itemize}
    \item \textbf{Differentiable Metasurface Inverse Design Framework:} We propose an approach to inverse-design metasurfaces with physical realizability guarantees. By formulating metasurface design as a differentiable optimization problem within a deep learning pipeline, we overcome limitations of traditional electromagnetic solvers and enable gradient-based optimization of complex geometric structures.
    
    \item \textbf{Partial Crystallization as Compressive Filters:} We exploit optical properties of \gls{gsst} in its intermediate crystallization states to create a physically realizable encoding mechanism for compressive sensing. By adjusting crystallization degree across 11 distinct levels, our approach generates a rich array of spectral response functions that effectively compress high-dimensional \gls{swir} signals without requiring bulky optical components.
    
    \item \textbf{Physics-aware Neural Simulator with Shape Regularization:} We develop a neural simulator trained on over $320{,}000$ metasurface configurations that accurately predicts spectral responses. The soft shape regularization technique that encourages all metasurface designs remain manufacturable throughout optimization, addressing a critical challenge in computational metasurface design.
    
    \item \textbf{End-to-End Co-Design with Performance Gains:} Jointly optimizing metasurface geometry and reconstruction network yields a \SI{7.6}{dB} PSNR enhancement over unoptimized designs. Our framework improves reconstruction fidelity while enhancing the condition number of the measurement matrix, providing improved stability and noise resilience.
\end{itemize}

Our approach integrates computational optics, deep learning, and materials science to advance metasurface inverse design. The framework offers a generalizable approach for optical elements with specific spectral responses, showing consistent improvements across metrics and robust performance under noise conditions.

% combined_optical_properties

\begin{figure*}[!h]
\centering
\includegraphics[width=2\columnwidth]{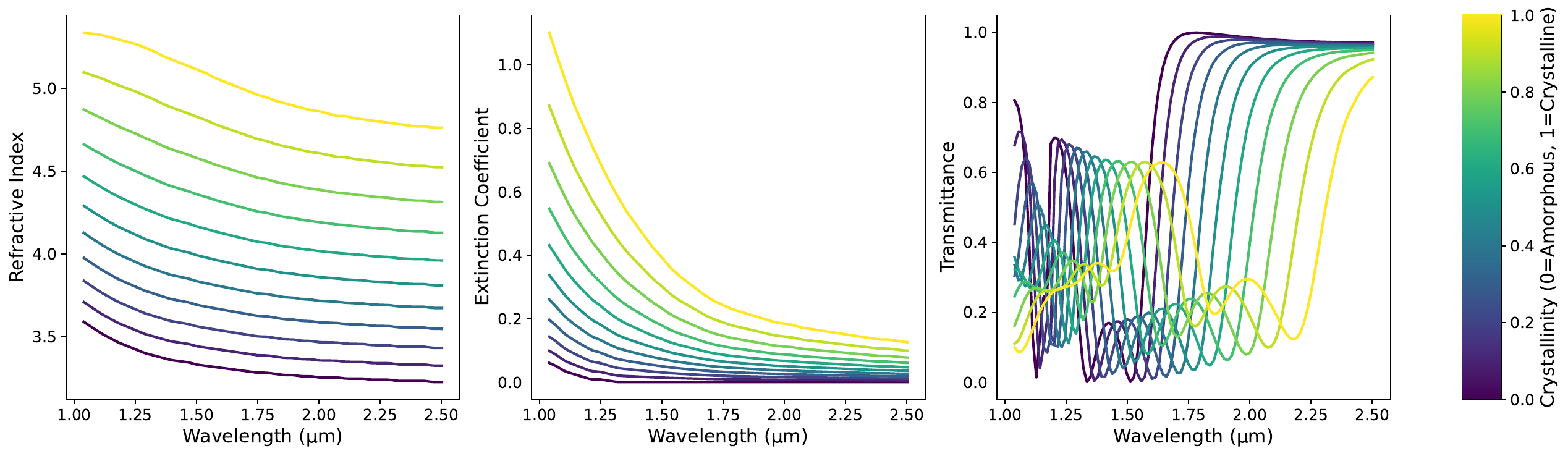}
\caption{Optical properties of GSST at varying crystallization levels. Left: Real part ($n$) of the refractive index obtained via effective medium theory for different crystallinity levels $C\in[0,1]$. Middle: Corresponding imaginary part ($k$) showing increased absorption at higher crystallization states. Right: Representative transmittance spectra of a GSST-based metasurface under 11 distinct crystallization levels, demonstrating how partial crystallization enables a diverse set of spectral filters for compressive measurements. The continuous tunability of crystallization fraction provides a rich space of physically realizable filter shapes across the SWIR spectrum.}
\label{fig:nk_partial}
\end{figure*}

%%%%%%%%%%%%%%%%%%%%%%%%%%%%%%%%%%%%%%%%%%%%%%%%%%%%%%%%%%%%%%%%%%%%%%%%
\section{Related Work}\label{sec:relatedwork}

\subsection{Inverse Design of Metasurfaces}
Designing metasurfaces for tailored optical functions is an inherently inverse problem in a high-dimensional, non-convex space. Traditional methods such as particle swarm optimization~\cite{wang2018particle} and genetic algorithms~\cite{back1993overview} are computationally intensive and non-differentiable, which hinders end-to-end integration with downstream tasks. Recent data-driven approaches leverage deep neural networks as efficient surrogates for electromagnetic solvers~\cite{wiecha2021deep, so2020deep, pestourie2020active}, though they often face challenges ensuring physical realizability. Our work builds on these advances by introducing a neural simulator trained on over 320,000 \gls{gsst} metasurface geometries with spectral responses across 11 crystallization states. Using a Transformer-based architecture~\cite{zhong2021spectral, lee2019set}, the simulator accurately maps geometric features to spectral behavior and, being fully differentiable, enables gradient-based optimization of both the metasurface and the spectral reconstruction network~\cite{molesky2018inverse}. Additionally, our novel soft shape regularization technique ensures that the optimized designs remain physically realizable~\cite{mansouree2020multifunctional}, addressing one of the key challenges in bridging computational design with practical fabrication constraints.

\subsection{Phase-Change Metasurfaces}
While the neural framework drives our computational design, physical implementation requires materials with specific optical properties. \gls{pcm}s such as \gls{gsst} exhibit a high refractive index contrast between amorphous and crystalline states and offer broadband transparency in the \gls{swir} region~\cite{zhang2019broadband, burtsev2022physical}. These materials enable non-volatile, reversible tuning without mechanical parts. Although PCMs have been applied to optical switching~\cite{rude2016ultrafast}, lens tuning~\cite{gu2023reconfigurable}, and holography, exploiting intermediate crystallization states beyond simple binary transitions remains underexplored. By discretizing the GSST crystallization fraction into 11 distinct levels—each serving as a unique spectral filter~\cite{tao2022shortwave, julian2020reversible}—our approach expands the design space for tunable filtering and simplifies the optical stack as illustrated in Fig.~\ref{fig:nk_partial}.

\subsection{Compressive Hyperspectral Imaging}
By leveraging the sparsity inherent in hyperspectral data, compressive sensing techniques recover high-dimensional spectral signals from significantly fewer measurements~\cite{gehm2007single, gerrits2018short}. Early implementations relied on dispersive elements and narrow-band tunable filters, yet these systems are often limited by calibration complexities and mechanical constraints. Metasurface-based compressive hyperspectral imaging provides an alternative to bulky spectrometers and mechanical filter wheels~\cite{goetz1985imaging, meng2020snapshot}. Our design-centric approach directly addresses these limitations by optimizing GSST-based metasurfaces as optical encoders, searching through diverse spectral response functions to find the candidate that are well-suited for the reconstruction task~\cite{li2022inverse}.

\subsection{Deep Learning for Spectral Reconstruction}
Recent progress in deep learning has significantly advanced the reconstruction of hyperspectral data from compressed measurements~\cite{zhang2022survey, huang2022spectral}. In contrast to traditional \gls{cs} algorithms that rely on sparsity and linear inversion, neural networks—especially those employing convolutional or transformer-based architectures~\cite{hong2021spectralformer, zhong2021spectral}—can directly learn complex spectral priors, offering improved reconstruction accuracy under noisy and variable conditions~\cite{qian2021hyperspectral}. We extend this paradigm using a two-stage reconstruction framework that first maps encoded measurements into a latent geometric representation before decoding them back into the spectral domain. By co-optimizing this reconstruction network with our inverse-designed \gls{gsst} metasurface encoder, we ensure that both the physical and computational components are mutually adapted, achieving state-of-the-art performance in spectral reconstruction~\cite{tseng2021neural, chang2019deep}.

% \newpage

% \lipsum[2-13]
%%%%%%%%%%%%%%%%%%%%%%%%%%%%%%%%%%%%%%%%%%%%%%%%%%%%%%%%%%%%%%
\section{Methods}
We present a unified, end‑to‑end differentiable framework for the co‑design of GSST‑based metasurface encoders and neural decoders in compressive hyperspectral reconstruction. Central to our methodology is a neural surrogate—trained on RCWA simulations across 11 discretized crystallization states—that functions as a physically grounded forward model, orchestrating a diverse ensemble of spectral filters. By integrating this surrogate within the reconstruction network, we jointly refine metasurface geometries and decoder parameters, yielding harmonized optical and computational components that achieve markedly improved spectral reconstruction fidelity.

%--------------------------------------------------------------------
\subsection{Forward Model for Spectral Encoding}\label{sec:forwardmodel}
\subsubsection{Phase-Change Tunability of \gls{gsst} Metasurfaces}\label{sec:emt}

GSST exhibits continuously tunable optical properties via partial crystallization~\cite{zhang2021electrically}. By varying the crystallization fraction $C\in(0,1)$, its effective permittivity $\epsilon(\lambda,C)$ follows the effective medium theory~\cite{xu2021nonvolatile}:
\begin{equation}
\frac{\epsilon(\lambda,C)-1}{\epsilon(\lambda,C)+2}
= C\,\frac{\epsilon_{c}(\lambda)-1}{\epsilon_{c}(\lambda)+2}
+ (1-C)\,\frac{\epsilon_{a}(\lambda)-1}{\epsilon_{a}(\lambda)+2},
\end{equation}
where $\epsilon_{c}(\lambda)$ and $\epsilon_{a}(\lambda)$ denote the permittivities of fully crystalline and amorphous GSST, respectively. From
\begin{equation}
\epsilon(\lambda,C)=\bigl[n(\lambda,C)+i\,k(\lambda,C)\bigr]^2,
\end{equation}
where $n(\lambda,C)$ is the refractive index and $k(\lambda,C)$ is the extinction coefficient, both vary continuously with $C$. When such GSST films are patterned into metasurfaces, these variations in $n$ and $k$ directly translate into changes in the spectral transmittance, effectively creating a tunable optical filter. We discretize $C$ into 11 levels $\{0.0,0.1,\dots,1.0\}$, generating a set of distinct spectral responses that form our measurement matrix $\mathbf{\Phi}$. Partial crystallization is controlled via electrical pulse intensity or width~\cite{abdollahramezani2022electrically}, enabling versatile spectral filtering without any mechanical components.

\subsubsection{Compressive Spectral Measurement Model}
We discretize the \gls{swir} band (1.0--\SI{2.5}{\um}) into $N=100$ spectral channels. For a hyperspectral datacube $\mathbf{X}\in \mathbb{R}^{H\times W\times N}$, we configure the metasurface in $M=11$ distinct crystallization states to obtain compressed measurements. For each pixel $(h,w)$, the recorded intensity in state $i$ is:
\begin{equation}
    y_{h,w}^{(i)} = \int_{\lambda_1}^{\lambda_2} x_{h,w}(\lambda)\,\mathbf{\Phi}_i(\lambda)\,d\lambda + n_{h,w}^{(i)},
\end{equation}
where $\mathbf{\Phi}_i(\lambda)$ is the filter response at state $C_i$, and $n_{h,w}^{(i)}$ denotes measurement noise. In discrete form:
\begin{equation}
    \mathbf{y}_{h,w} = \mathbf{\Phi}\,\mathbf{x}_{h,w} + \mathbf{n}_{h,w},
\end{equation}
with $\mathbf{\Phi} \in \mathbb{R}^{M\times N}$ encoding filter responses for each state. Since $M \ll N$, reconstruction requires compressive sensing methods. The full compressed measurement datacube is $\mathbf{Y}\in \mathbb{R}^{H\times W\times M}$.

\begin{figure}[h]
  \centering
  \includegraphics[width=0.95\linewidth]{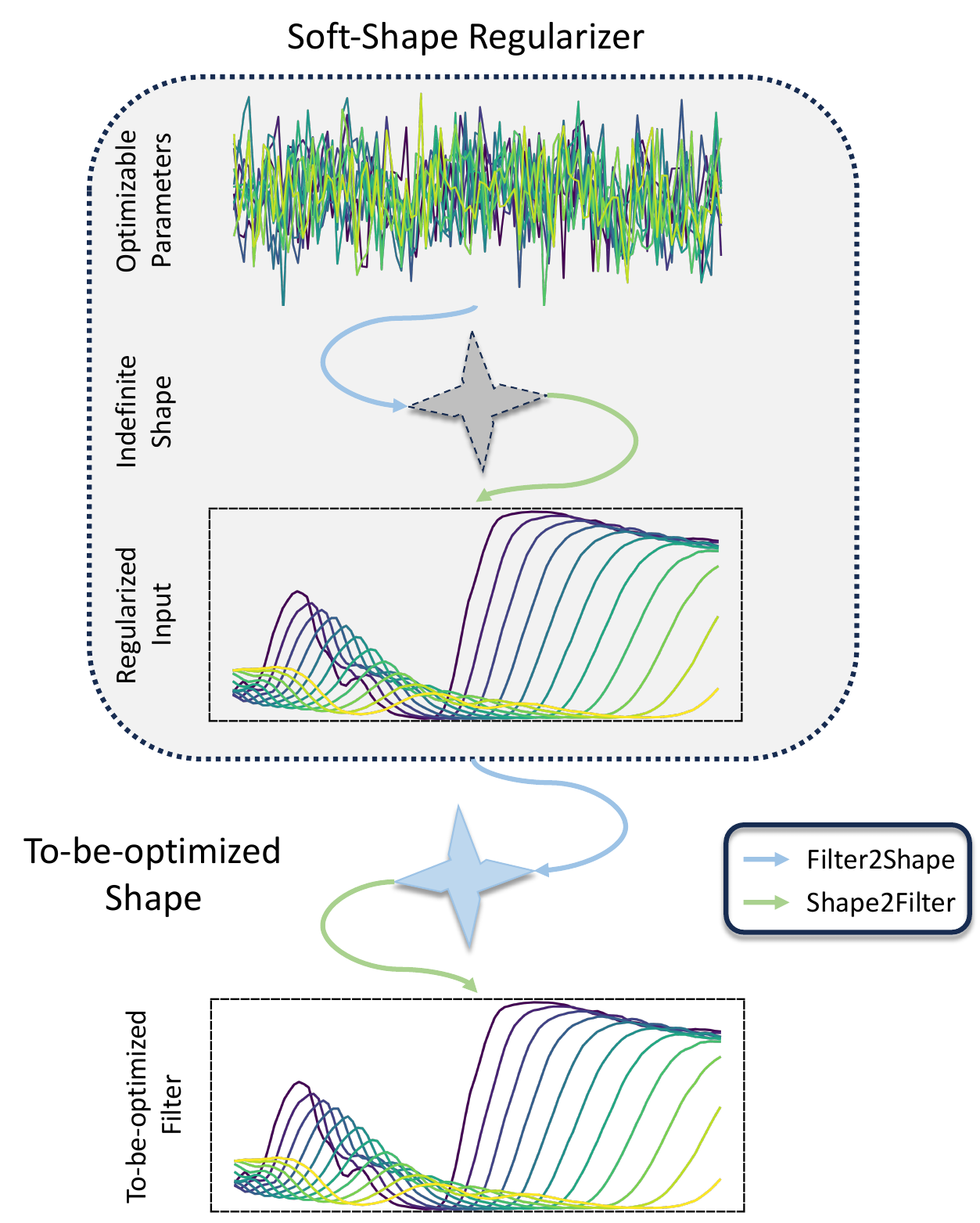}
  \caption{
    Soft shape regularization pipeline for optimization. To ensure physical realizability during gradient-based optimization, we employ a multi-stage transformation pathway: filter parameters are first mapped to candidate shapes via a learned filter2shape network, then validated through a shape2filter network, before re-mapping to shapes.
  }
  \label{fig:soft_shape_regularizer}
\end{figure}

%--------------------------------------------------------------------
\subsection{High-fidelity Neural Simulator}\label{sec:neural_simulator}
We develop a Transformer-based neural simulator $\mathcal{N}$ mapping 2D metasurface geometry to spectral responses across 11 crystallization states. Trained on 320,000+ configurations, our simulator replaces computationally intensive electromagnetic solvers with a differentiable approximation while constraining solutions to physically realizable designs.

\subsubsection{Dataset Preparation}
We generate 320,000+ diverse geometries with physical symmetry by sampling up to four vertices in the \gls{Q1} and mirroring them to enforce four-fold rotational symmetry, thereby ensuring polarization-independent spectral transmittance. Vertices are sampled at equal angular intervals to enhance manufacturability. For each geometry, we compute spectral transmittance under all 11 crystallization levels using S4, a \gls{rcwa} simulation software~\cite{liu2012s4}.

\subsubsection{Shape Regularization for Stable Optimization}
A key challenge in metasurface optimization is maintaining physical realizability during gradient updates~\cite{jiang2019global, molesky2018inverse}. We implement two key mechanisms: (1) a probabilistic vertex dependency where $p(v_i) = 1$ for $i=0$ and $p(v_i) = \sigma(l_i) \cdot p(v_{i-1})$ for $i>0$,, with progressively decreasing logit priors to encourage simpler shapes with fewer vertices; and (2) a soft shape regularization pipeline (Fig.~\ref{fig:soft_shape_regularizer}) using a differentiable filter$\rightarrow$shape$\rightarrow$filter$\rightarrow$shape cycle. First, a shape2filter model is trained to predict spectral responses from geometries. Second, a filter2shape model learns the inverse mapping. Finally, we freeze the shape2filter model and fine-tune the filter2shape model by minimizing the spectral error when connecting them in tandem.
%This approach maps spectral filters to plausible geometries, validates them through a frozen shape2filter network, and re-maps to shape parameters, keeping updates within physically realizable manifolds.

\subsubsection{Network Architecture}
Our Transformer architecture~\cite{dosovitskiy2020image} takes an input tensor $(4 \times 3)$ representing \gls{Q1} vertices with presence indicators and coordinates. The output is an $(M \times N)$ matrix of transmittance spectra across crystallization levels. Training minimizes \gls{mse} between predicted and true spectral responses:
\begin{equation}
    \mathcal{L}_{\mathcal{N}} = \frac{1}{MN} \sum_{i=1}^{M} \sum_{j=1}^{N} (\mathbf{\Phi}_{ij}^{\text{pred}} - \mathbf{\Phi}_{ij}^{\text{true}})^2.
\end{equation}

\subsection{Hyperspectral Reconstruction Network}\label{sec:reconstruction_network}
Given the forward model described in Section~\ref{sec:forwardmodel}, the goal of this module is to reconstruct the high-dimensional hyperspectral datacube $\mathbf{X}\in \mathbb{R}^{H\times W\times N}$ from the compressed measurement datacube $\mathbf{Y}\in \mathbb{R}^{H\times W\times M}$, where $M=11$ corresponds to the number of crystallization states of the GSST-based metasurface. To this end, we adopt a modified version of the Adaptive Weighted Attention Network architecture~\cite{li2020adaptive}, originally proposed for RGB-to-HSI spectral reconstruction, and adapt it to our unique spectral encoding scenario.

\subsubsection{Network Architecture}

Our reconstruction model, denoted by $\mathcal{D}_\theta$ (with $\theta$ representing its learnable weights), uses two stacked Dual Residual Attention Blocks (DRABs) as its core feature extraction units. Each DRAB integrates both long and short residual connections to preserve low-frequency information and ensure stable gradient propagation. The input tensor has a shape of $(H, W, M)$, representing the $11$-channel encoded measurements from the metasurface, as defined in Section~\ref{sec:forwardmodel}.

% Each DRAB module includes:
% \begin{itemize}
%   \item A pair of residual branches (5×5 and 3×3 convolutions) with dual skip‑connections and PReLU activations, forming the dual residual learning.
%   \item An \textbf{Adaptive Weighted Channel Attention} sub‑module that adaptively recalibrates channel responses via learned spatial weights.
% \end{itemize}
Additionally, after stacking all DRABs, a \textbf{Patch‑level Second‑order Non‑local} module is appended at the network tail to split feature maps into patches and perform second‑order non‑local operations for capturing long‑range spatial dependencies.

The output of $\mathcal{D}_\theta$ is a hyperspectral datacube of shape $(H, W, N)$, where $N$ is the number of spectral bands.

\subsubsection{Joint Optimization with Metasurface Filter}

In our joint optimization framework, both the metasurface shape parameters $S$ and the decoder network parameters $\theta$ are optimized simultaneously. Specifically, we seek to minimize the reconstruction error between the ground truth hyperspectral datacube $\mathbf{X}$ and the decoder output $\mathcal{D}_\theta\bigl[\mathcal{N}(S) \cdot \mathbf{X}\bigr]$, where $\mathcal{N}$ is the neural simulator mapping $S$ to the corresponding spectral response matrix, and $\mathcal{D}_\theta$ is the reconstruction network parameterized by $\theta$. We solve
\begin{equation}
(\hat{S}, \hat{\theta}) = \arg\min_{S,\theta} \; \bigl\|\mathbf{X} - \mathcal{D}_\theta\bigl[\mathcal{N}(S) \cdot \mathbf{X}\bigr]\bigr\|_{2}^2.
\end{equation}

By simultaneously learning $S$ and $\theta$, the metasurface encoder and the decoder are co-adapted for improved hyperspectral reconstruction.

\begin{figure*}[!h]
  \centering
  \includegraphics[width=\textwidth]{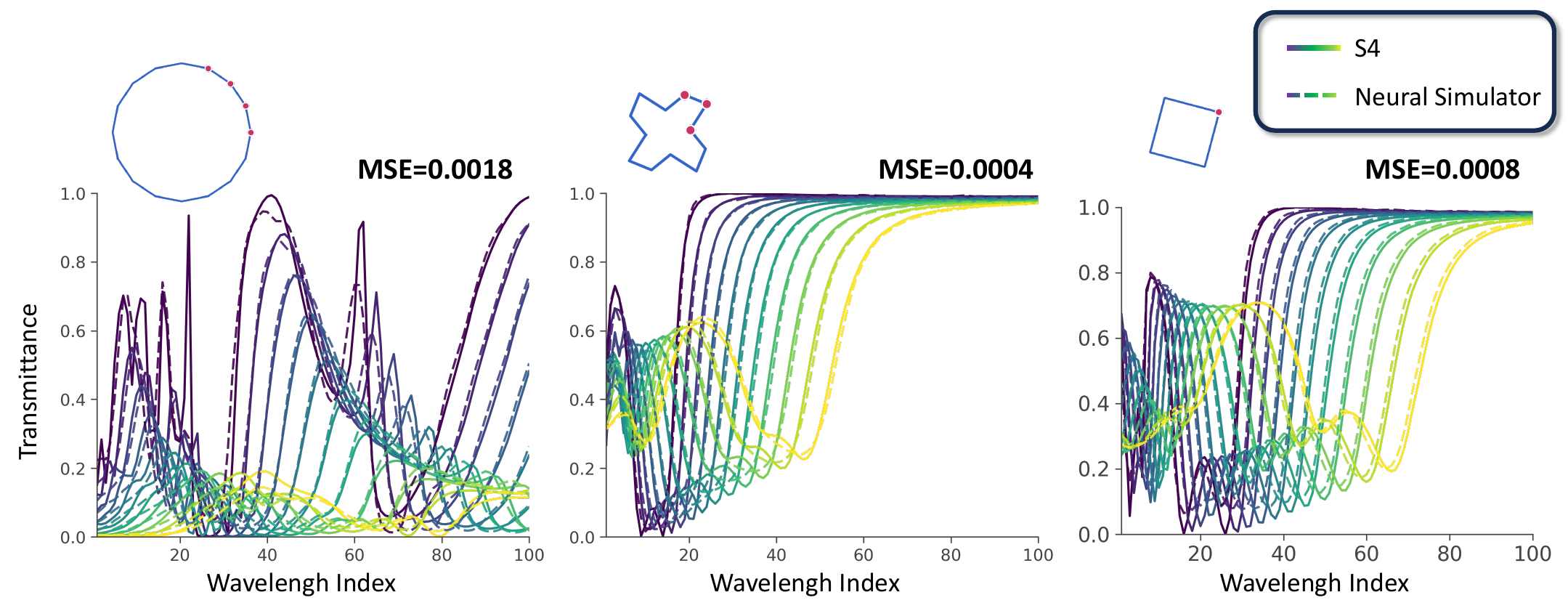}
  \caption{
  Representative examples of metasurface geometries (insets) and their corresponding spectral responses predicted by S4 and our neural simulator across 11 GSST crystallization states. The left panel shows the spectral response of a circular geometry (MSE=$0.0018$), the middle panel presents a concave plus-like structure (MSE=$0.0004$), and the right panel displays a square configuration (MSE=$0.0008$).
  }
  \label{fig:neural_sim_verification}
\end{figure*}

\begin{figure*}[!h]
  \centering
  \includegraphics[width=\textwidth]{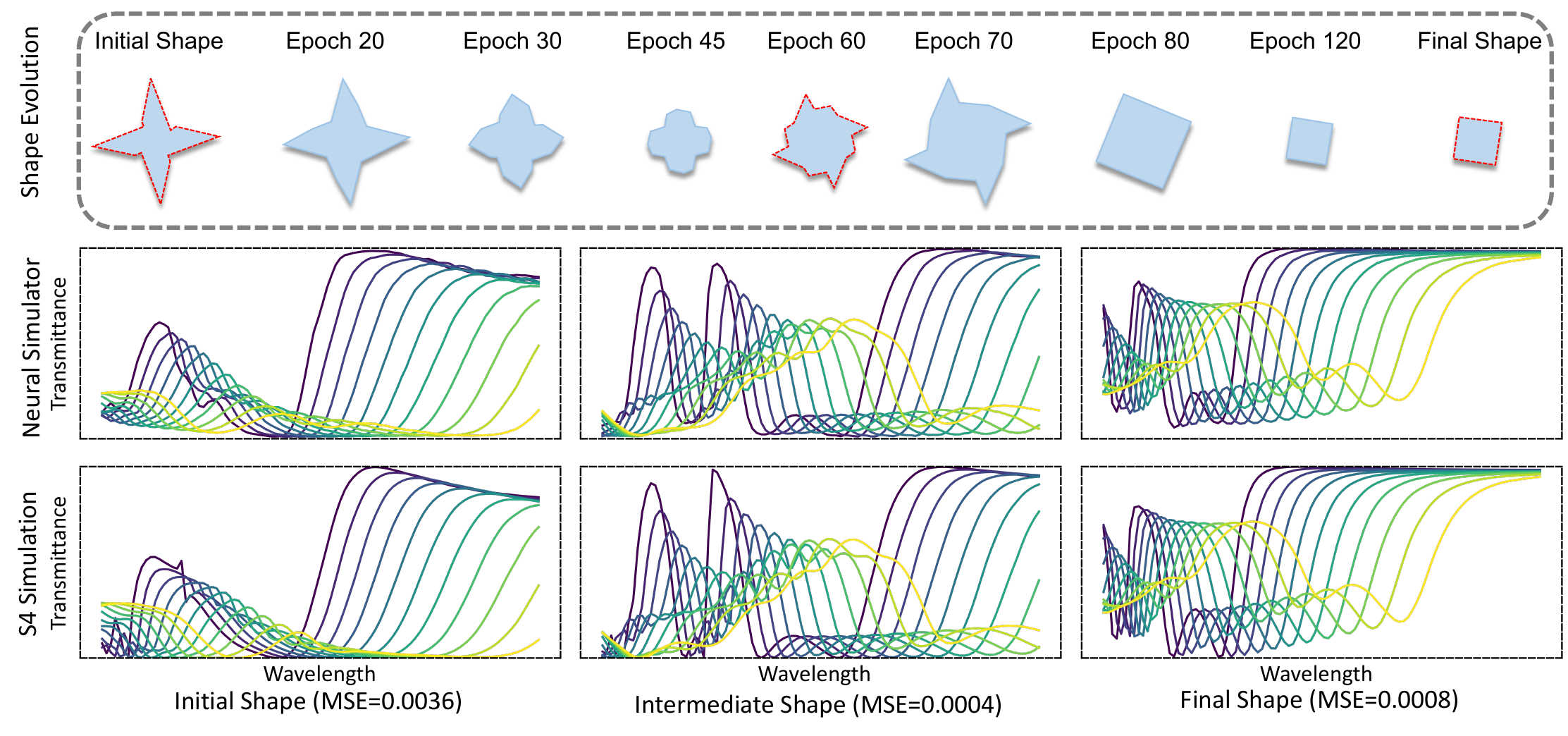}
    \caption{
    Evolution of the GSST metasurface geometry and its spectral response during joint optimization. \textbf{Top row:} Representative snapshots of the metasurface shapes at selected epochs over a $250$-epoch training process, with dashed boundaries highlighting three key shapes: the initial geometry (left), an intermediate shape at epoch $60$ (middle), and the final optimized structure (right). \textbf{Second and third rows:} Each column shows the spectral response corresponding to its respective dashed-outlined shape above. The first column displays spectra for the initial shape (MSE=$0.0036$), the second column for the intermediate shape (MSE=$0.0004$), and the third column for the final optimized shape (MSE=$0.0008$).
    }
  \label{fig:gsst_evolution}
\end{figure*}

\begin{figure*}[!h]
  \centering
  \includegraphics[width=1\textwidth]{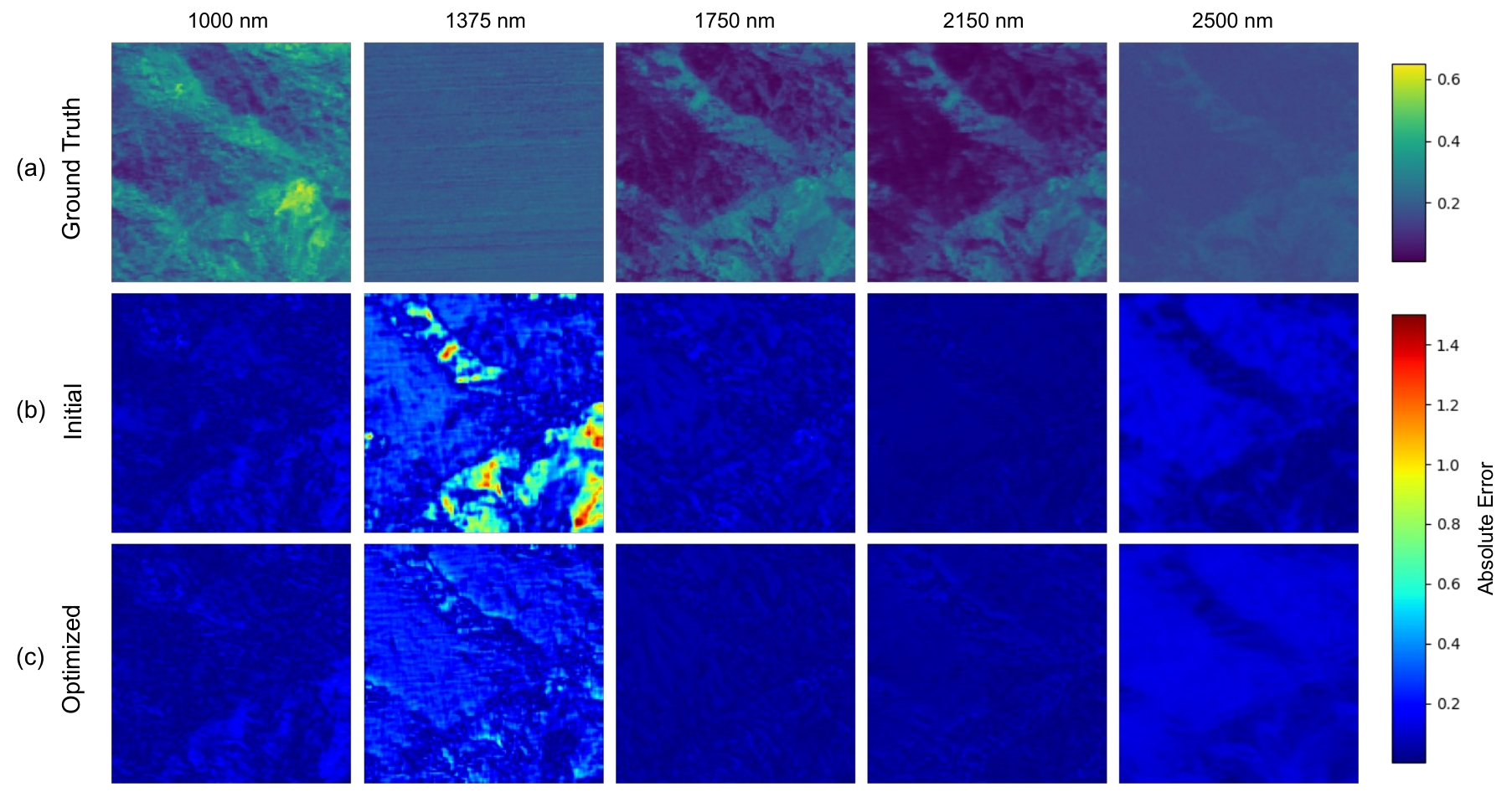}
  \caption{
  A randomly selected sample from the test set is shown. The top row (a) displays the ground truth images across five spectral bands (wavelengths ranging from \SI{1000}{nm} to \SI{2500}{nm}). The middle row (b) presents the absolute error maps for the initial filter, and the bottom row (c) shows the absolute error maps for the optimized filter. The right-hand colorbars indicate the corresponding data values for the ground truth and absolute errors, respectively.
  }
  \label{fig:bandwise_error_comparison}
\end{figure*}

\begin{figure*}[!h]
  \centering
  \includegraphics[width=\textwidth]{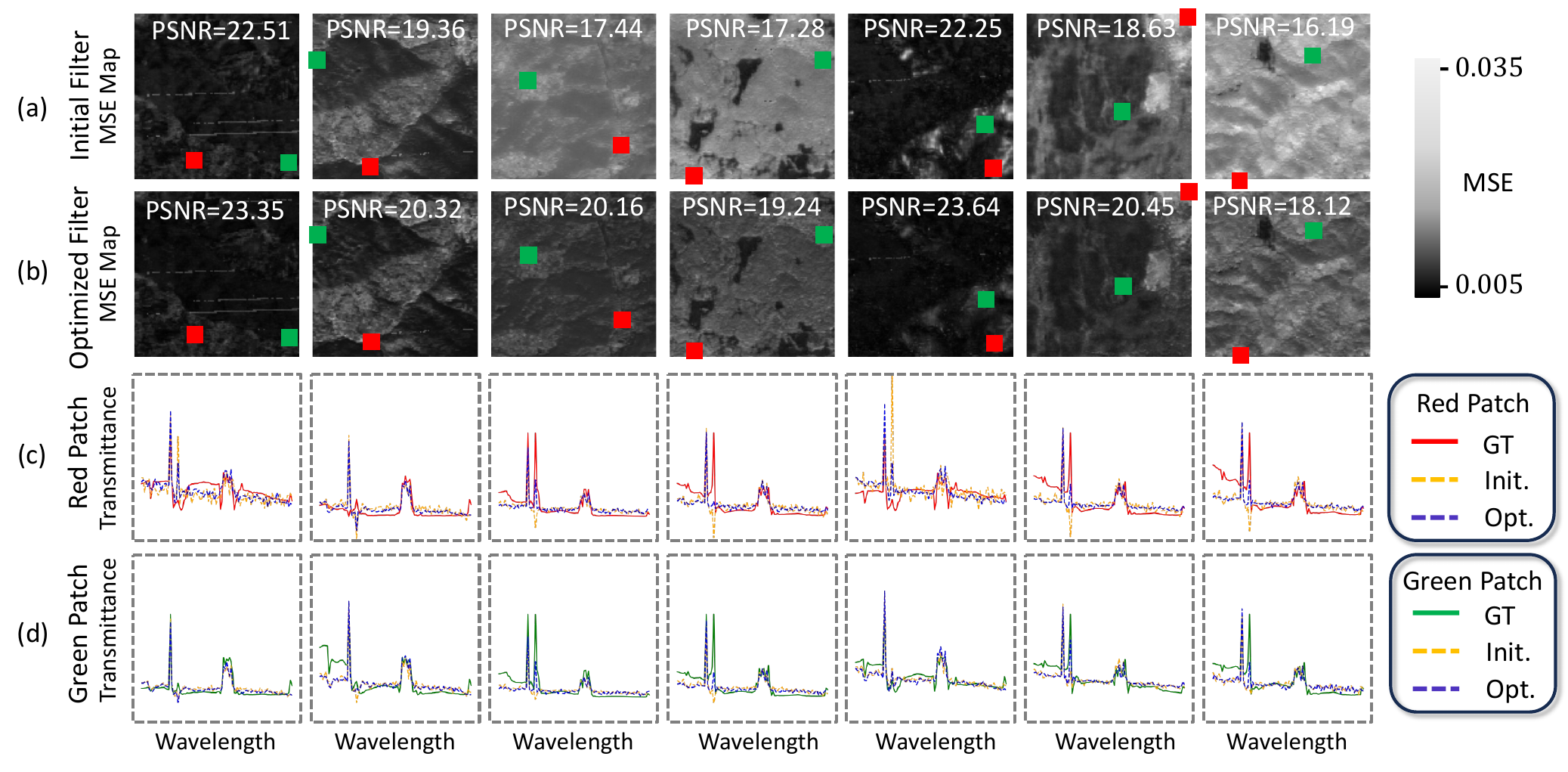}
  \caption{
    Comparison of hyperspectral reconstruction performance across seven test scenes.  
    (a) Spatial distributions of reconstruction \gls{mse} using the initial (unoptimized) filters. Each heatmap includes the corresponding \gls{psnr} value, with red and green points indicating two representative sampling locations.  
    (b) \gls{mse} heatmaps using the optimized filters, along with updated \gls{psnr} values.
    (c-d) Spectral reflectance comparisons at the red and green points, respectively, displaying ground truth and reconstructions from both initial and optimized filters.
  }
  \label{fig:spectral_point_comparison}
\end{figure*}

\begin{figure}[!h]
    \centering
    \includegraphics[width=0.9\linewidth]{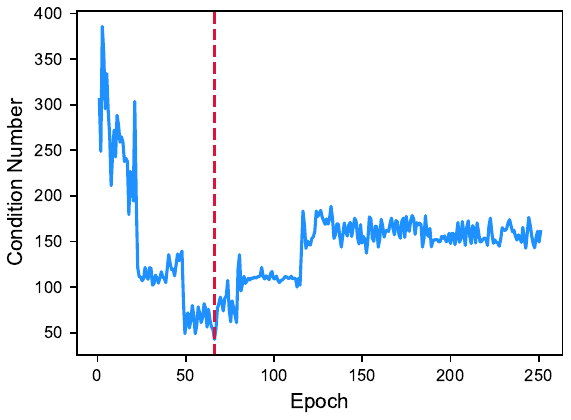}
    \caption{
    % Evolution of the condition number of the response matrix $\Phi$ during joint optimization. The condition number steadily decreases across training epochs, indicating improved numerical stability and better-conditioned spectral encodings.
    Evolution of the condition number of $\mathbf{\Phi}$ during joint optimization. The condition number initially decreases, reaching a minimum of 46.83 at epoch 70, before gradually increasing during later stages of training. 
    }
    \label{fig:cond_number}
\end{figure}

% \newpage

%%%%%%%%%%%%%%%%%%%%%%%%%%%%%%%%%%%%%%%%%%%%%%%%%%%%%%%%%%%%%%%%%%%%%%%%
\section{Experimental Results}\label{sec:experiments}
This section evaluates our proposed framework. We first describe our dataset (\textbf{3.1}), verify our neural simulator's accuracy (\textbf{3.2}), present joint optimization results (\textbf{3.3}), analyze matrix condition numbers (\textbf{3.4}), and examine noise robustness (\textbf{3.5}).

\subsection{Dataset and Spectral Preprocessing} \label{sec:dataset}
Our experiments use the \gls{aviris} dataset, extracting the \gls{swir} spectral region from $1.0$ to \SI{2.5}{\um}. We discretize this range into $N = 100$ uniform spectral channels. We preserve all spectral bands, including those with high atmospheric absorption or sensor noise (referred to as ``bad bands''), allowing our model to learn directly from the raw spectral characteristics.

\subsection{Verification of Neural Simulator}
Our neural simulator maps GSST metasurface shapes to spectral responses across 11 crystallization states, achieving a test set \gls{mse} of $0.0016$, showcasing the high accuracy of the neural simulator. To further validate the performance, we compare the ground truth calculated by S4 and the neural simulator's predictions for the spectral response functions of three distinct GSST-based metasurface shapes across all 11 crystallization states. As shown in the three comparison panels of Fig.~\ref{fig:neural_sim_verification}, the results highlight the simulator's capability to accurately capture the spectral behavior, and map geometrical features to their multi-state optical responses.

\subsection{Joint Optimization of Metasurface Encoder and Network Decoder}
To simulate a realistic spectral measurement environment, we add zero-mean Gaussian noise to the encoded measurements during training. Specifically, we compute the noise standard deviation based on a target \gls{snr} ranging from \SI{10}{dB} to \SI{40}{dB}, with the \gls{snr} uniformly sampled throughout training to ensure robustness under varying noise conditions.

Our training strategy consists of two stages. In the first stage, we randomly initialize the GSST metasurface shape and jointly optimize both the network decoder and the metasurface encoder (i.e., the shape and its associated spectral responses predicted by our neural simulator) over $250$ epochs. During this joint optimization, the metasurface shapes evolve from random initial geometries to more refined structures that yield increasingly distinct and well-behaved spectral responses. Fig.~\ref{fig:gsst_evolution} shows representative snapshots of this evolution, where the top row illustrates the progression of the metasurface geometries and the spectra from our neural simulator with the corresponding S4 simulation results for the initial, intermediate, and final shapes. The close agreement between the neural simulator predictions and S4 simulation results validates our model's accuracy, while the evolution in shape demonstrates the enhanced filtering capabilities achieved by joint optimization.

In the second stage, we further evaluate the impact of the optimized metasurface design. Specifically, both the initial and the optimized shapes are separately fed into the S4 model to obtain their corresponding \gls{rcwa}-simulated spectral responses. These responses are then fixed and the decoder is retrained independently for each case. This two-stage training strategy allows us to isolate and quantify the benefits of the optimized metasurface design on the overall hyperspectral reconstruction performance.

For testing, we evaluate the system under the measurement environment with an \gls{snr} of \SI{40}{dB}. Quantitative results shows that optimized filters significantly outperform unoptimized ones, with improvements of \SI{7.6}{dB} in \gls{psnr} (from \SI{15.10}{dB} to \SI{22.76}{dB}), reduced \gls{sam} from \SI{0.2076}{rad} to \SI{0.1623}{rad}, and 83\% reduction in \gls{mse} (from $0.0309$ to $0.0052$). These significant improvements across all three metrics underscore the benefits of integrating metasurface design into the learning loop.

We further analyze the spectral fidelity by inspecting five wavelength bands from the reconstructed hyperspectral datacubes. As shown in Fig.~\ref{fig:bandwise_error_comparison}, we visualize the ground truth reflectance for each band, alongside the corresponding absolute error maps for both the initial and the optimized filter reconstructions. These band-wise comparisons confirm that the optimized filters consistently yield lower spectral reconstruction errors, further validating the benefit of our joint learning framework.

To provide a evaluation of spectral reconstruction accuracy, we conduct a multi-faceted analysis across seven test samples. As shown in Fig.~\ref{fig:spectral_point_comparison}, the first and second rows present heatmaps of reconstruction \gls{mse} using unoptimized and jointly optimized filters, respectively, revealing significant error reduction across all scenes after optimization.

For each test sample, we randomly select two representative locations (marked on the \gls{mse} heatmaps) to analyze spectral fidelity. The third and fourth rows compare the ground truth reflectance spectra against reconstructions from both filter configurations at these locations. This visualization demonstrates that: (1) the optimized filters consistently yield lower \gls{mse} across entire scenes, and (2) the reconstructed spectra from optimized filters show markedly better alignment with ground truth at both sampled points, particularly in preserving characteristic spectral features.

The systematic improvement observed in both spatial error distributions and point-wise spectral reconstructions provides strong evidence for the superiority of our joint optimization approach. This quantitative and qualitative enhancement is maintained across diverse scenes, confirming the robustness of our encoder-decoder framework.

\subsection{Condition Number Analysis of the Response Matrix}
To further evaluate the effect of our joint optimization strategy, we analyze the condition number of the response matrix $\mathbf{\Phi}$ throughout training. A lower condition number generally indicates that the measurement matrix is better conditioned, which can improve reconstruction robustness and reduce noise amplification.

The condition number of response matrix $\mathbf{\Phi}$ 
%exhibits a non-monotonic trend during training (Fig.~\ref{fig:cond_number}). It 
initially decreases to a minimum of $43.65$ around epoch 60 before gradually increasing. At this minimum point, we observe notable improvements in reconstruction metrics (\SI{3.21}{dB} \gls{psnr} gain, 0.05 \gls{sam} reduction, 0.015 \gls{mse} reduction) over the initial design.

Interestingly, reconstruction performance continues to improve beyond this point, suggesting that while a well-conditioned sensing matrix contributes to improved reconstruction, the compatibility between encoder's spectral characteristics and decoder's learned priors also significantly impacts overall system performance.

\subsection{Performance Under Different Noise Levels}
During training, the measurement noise is uniformly sampled over an \gls{snr} range of $10$ to \SI{30}{dB} to simulate realistic noise variations. Although earlier results were presented for a measurement environment with an \gls{snr} of \SI{30}{dB}, here we further demonstrate the robustness of our approach by evaluating its performance in environments with \gls{snr} values of $10$, $20$, and \SI{30}{dB}. Table~\ref{table:noise_performance} summarizes the performance metrics—\gls{psnr}, \gls{sam}, and \gls{mse}—comparing the hyperspectral reconstruction using unoptimized filters with that using optimized filters. As shown by the results, the optimized filters consistently yield higher \gls{psnr}, lower \gls{sam}, and lower \gls{mse} across all these noise conditions, underscoring the strong generalization capability of our joint optimization framework to diverse measurement environments.

% Mountain Dataset
\begin{table}[h]
  \centering
  \caption{Reconstruction performance at three noise levels.}
  \label{table:noise_performance}
  \begin{tabular}{cccccc}
    \toprule
    \multirow{2}{*}{\gls{snr}} & \multirow{2}{*}{Method} & \textbf{\gls{psnr}} $\uparrow$ & \textbf{SAM} $\downarrow$ & \textbf{MSE} $\downarrow$ \\
    \cmidrule{3-5}
    & & (dB) & (rad) & \\
    \midrule
    \multirow{2}{*}{10} 
      & Unoptimized & 15.10 & 0.2066 & 0.0308 \\
      & Optimized   & \textbf{22.71} & \textbf{0.1658} & \textbf{0.0053} \\
    \midrule
    \multirow{2}{*}{20} 
      & Unoptimized & 15.10 & 0.2088 & 0.0309 \\
      & Optimized   & \textbf{22.69} & \textbf{0.1642} & \textbf{0.0053} \\
    \midrule
    \multirow{2}{*}{30} 
      & Unoptimized & 15.09 & 0.2081 & 0.0309 \\
      & Optimized   & \textbf{22.74} & \textbf{0.1640} & \textbf{0.0053} \\
    \bottomrule
  \end{tabular}
\end{table}

% % Forest Dataset
% \begin{table}[t]
%   \centering
%   \caption{Reconstruction performance at three additional noise levels.}
%   \label{table:noise_performance}
%   \begin{tabular}{cccccc}
%     \toprule
%     \multirow{2}{*}{$SNR$} & \multirow{2}{*}{Method} & \textbf{\gls{psnr}} $\uparrow$ & \textbf{SAM} $\downarrow$ & \textbf{MSE} $\downarrow$ \\
%     \cmidrule{3-5}
%     & & (dB) & (rad) & \\
%     \midrule
%     \multirow{2}{*}{10} 
%       & Unoptimized & 34.18 & 0.068376 & 0.000382 \\
%       & Optimized   & \textbf{34.97} & \textbf{0.062757} & \textbf{0.000318} \\
%     \midrule
%     \multirow{2}{*}{20} 
%       & Unoptimized & 34.37 & 0.066426 & 0.000366 \\
%       & Optimized   & \textbf{35.46} & \textbf{0.060155} & \textbf{0.000284} \\
%     \midrule
%     \multirow{2}{*}{30} 
%       & Unoptimized & 34.38 & 0.067494 & 0.000365 \\
%       & Optimized   & \textbf{35.55} & \textbf{0.058921} & \textbf{0.000279} \\
%     \bottomrule
%   \end{tabular}
% \end{table}

\lipsum[2-13]

%%%%%%%%%%%%%%%%%%%%%%%%%%%%%%%%%%%%%%%%%%%%%%%%%%%%%%%%%%%%%%%%%%%%%%%%
\section{Discussion and Conclusion}\label{sec:discussion_conclusion}
In this work, we proposed a data-driven framework for hyperspectral imaging based on phase-change metasurfaces made from \gls{gsst}. By introducing a differentiable neural simulator that maps metasurface geometries to their spectral responses across 11 crystallization states, we replaced traditional electromagnetic solvers (e.g., \gls{fdtd}, \gls{rcwa}) with an efficient approximation that enables joint optimization of the optical encoder and the reconstruction network.

Our framework allows the metasurface filters to be co-optimized with a deep hyperspectral reconstruction network in an end-to-end manner. Experimental results on the \gls{aviris} dataset demonstrate that the jointly optimized filters significantly outperform unoptimized filters, as evidenced by improvements in \gls{psnr}, \gls{sam}, and \gls{mse}. Quantitative tables and visual comparisons confirm enhanced spectral fidelity and robustness to measurement noise.

In conclusion, our approach demonstrates that jointly learning metasurface hardware and computational reconstruction models from data is not only feasible but also highly effective. This work opens new possibilities for developing compact, tunable, and intelligent hyperspectral imaging systems powered by phase-change materials and neural networks.

\ifpeerreview \else
\section*{Acknowledgments}
We gratefully acknowledge discussions with the metasurface and computational imaging communities at The University of Hong Kong. This work was partially supported by xxx.
\fi

% \newpage

\bibliographystyle{IEEEtran}
\bibliography{references}

% \newpage

\ifpeerreview \else
%%%% For the camera-ready version, a short author biography is often included.
\begin{IEEEbiography}{Rongzhou Chen}
Biography text (camera-ready). 
\end{IEEEbiography}

\begin{IEEEbiographynophoto}{Haitao Nie}
Biography text (camera-ready).
\end{IEEEbiographynophoto}

\begin{IEEEbiographynophoto}{Shuo Zhu}
Biography text (camera-ready).
\end{IEEEbiographynophoto}

\begin{IEEEbiographynophoto}{Edmund Y. Lam}
Biography text (camera-ready).
\end{IEEEbiographynophoto}
\fi

\end{document}